# Rotating compact objects with magnetic fields

Anshu Gupta[*], Amruta Mishra[†], Hiranmaya Mishra, A.R. Prasanna

*Theory Division, Physical Research Laboratory, Navrangpura, Ahmedabad -380 009, India*

## Abstract

We consider here the structure of rotating compact objects endowed with a magnetic field in general relativity as models of pulsars. We discuss first the structure of rotating stars in the framework of Hartle taking different realistic equations of state and study their effects on bulk properties of the star. We consider the possibilty of rotating stars with a quark matter core. We further analyse the structure of the magnetic field in the interior of the star as affected by different equations of state as well as due to rotation.

Typeset using REVTEX

[*]E-mail: anshu@prl.ernet.in

[†]E-mail: am@prl.ernet.in

# I. INTRODUCTION

The study of structure and dynamics of neutron stars is an interesting and challenging theoretical problem. The system here is quite complicated. The large density in the interior of such stars can lead to a possibility of melting the baryons to form a deconfined quark core [1]. Observations of rotating neutron stars with a period in the millisecond range has renewed the interest in the study of the influence and implications of rotation on the bulk properties of such objects [2]. Further the neutron stars are also known to possess strong magnetic fields. Their polar field strengths are deduced from the observed spin slowdown of pulsars via the magnetic dipole breaking model with most of the young pulsars having a surface field, $\mathbf{B} \approx 10^{12}$ gauss [4]. Each one of the above facets of neutron stars poses interesting theoretical challenges. The fascinating possibility of a deconfined phase of matter in the interior of neutron stars has stimulated the work of many authors proposing different signals of existence of such a core of quark matter. These studies mostly have been done for considering nonrotating stars. The treatment of massive compact rotating objects in the framework of general theory of relativity is a rather complicated task. However, for slow rotation this can be done in the framework of Hartle and Thorne [3]. Further, the magnetic field structure outside the neutron star has been studied in connection with pulsar emission mechnism [5]. In a nonrelativistic approach, the generation and evolution of internal magnetic fields have been studied recently [6]. A relativistic treatment to include structure of magnetic field in the interior of neutron stars by solving coupled Einstein Maxwell's equations has been done numerically in ref. [7]. It is therefore interesting to incorporate all these effects together in its totality to the extent possible and the present work is a step in that direction.

We shall consider rotating stars endowed with magnetic fields. The effects of rotation is taken into account in the framework of Hartle and Thorne where rotation effect is treated as a perturbation over the Schwartzchild solution and is valid naturally for slowly rotating stars [3,8]. The field structure shall be assumed to be poloidal and axisymmetric. The effect of rotation of the star on the magnetic field structure is included, but not the reverse, amounting to the fact that the magnetic field energy is not strong enough to affect the space time geometry. We shall also take into account the effects of different equations of state on the gross properties of the star and on the magnetic field in the interior.

We organise the paper as follows. In section II, we give a short description of treating rotating compact objects in general relativity in the approximations of Hartle and Thorne [3]. In section III, we discuss Maxwell's equations in curved space time with Hartle metric and set up the differential equations for electromagnetic fields. Here we obtain the boundary conditions on the dipole magnetic field structure taking corrections due to rotation upto $O(\Omega^2)$. Using these boundary conditions, we integrate them from the boundary to the interior of the star. In section IV, we discuss different equations of state and their effects on the bulk properties of the star along with the field structure in the interior. Section V summarizes the results and discussions.

## II. ROTATING NEUTRON STARS IN HARTLE THORNE APPROXIMATION

The description of rotating compact objects in the framework of general relativity is a complicated task. In what follows, we shall be using the method of Hartle and Thorne [3], appropriate for slowly rotating objects. Here a perturbative solution is formulated over the Schwarzschild metric for the nonrotating stars given as

$$ds^2 = -e^{\nu_0}dt^2 + (1 - 2GM/r)^{-1}dr^2 + r^2(d\theta^2 + sin^2\theta d\phi^2) \tag{1}$$

The metric function is expanded in terms of the star's rotational frequency, $\Omega$, the angular velocity of the star's surface relative to an observer at infinity and is given as

$$ds^2 = -e^{2\nu}dt^2 + e^{2\lambda}dr^2 + e^{2\mu}d\theta^2 + e^{2\psi}(d\phi - \omega dt)^2 \tag{2}$$

In the above $\omega$ is the angular velocity of the star's fluid in a local inertial frame and depends on the radial coordinate $r$. The difference $\bar{\omega} = \Omega - \omega$ is the angular velocity with which the fluid inside the star moves relative to a distant observer. The metric functions are expanded in multipoles around the Schwarzchild metric as

$$e^{2\nu} = e^{\nu_0} \left[1 + 2(h_0 + h_2 P_2)\right], \tag{3a}$$

$$e^{2\lambda} = (1 - 2GM/r)^{-1} \left[1 + \frac{2G}{r}(m_0 + m_2 P_2)/\left(1 - \frac{2GM}{r}\right)\right], \tag{3b}$$

$$e^{2\mu} = r^2 \left[1 + 2(v_2 - h_2) P_2\right], \tag{3c}$$

$$e^{2\psi} = r^2 \sin^2\theta \left[1 + 2(v_2 - h_2) P_2\right]. \tag{3d}$$

In the above, the perturbation functions $m_0$, $m_2$, $h_0$, $h_2$ and $v_2$ are to be calculated from Einstein's field equations. Under the assumption that due to rotation the star distorts and affects the energy and pressure of the rotating fluid by $\Delta\epsilon$ and $\Delta p$ respectively i.e. the source term of Einstein equation $G_{\mu\nu} = 8\pi G T_{\mu\nu}$ changes by $\Delta T_{\mu\nu}$, one can perform a multipole expansion of these quantities as

$$\Delta p = (\epsilon + p)\left[p_0^* + p_2^* P_2\right] \tag{4a}$$

$$\Delta \epsilon = (\epsilon + p)\frac{d\epsilon}{dp}\left[p_0^* + p_2^* P_2\right] \tag{4b}$$

Finally, the quantities $\nu_0(r)$, $M(r)$ denoting the metric function and the mass profile of the nonrotating star, are obtained by solving the Tolman Oppenheimmer Volkhoff (TOV) equations

$$\frac{dp}{dr} = -\frac{G\left[\epsilon + p\right]\left[M + 4\pi r^3 p\right]}{r^3 \quad (r - 2GM)} \tag{5a}$$

$$\frac{dM}{dr} = 4\pi r^2 \epsilon \tag{5b}$$

$$\frac{d\nu_0}{dr} = -\frac{2}{(\epsilon + p)}\frac{dp}{dr} \tag{5c}$$

In what follows, we shall be limiting rotational deformations to monopole approximation only. The differential equations for the perturbation functions are given as

$$\frac{dm_0}{dr} = 4\pi r^2 \frac{d\epsilon}{dp}(\epsilon+p)p_0^* + \frac{1}{12G}j^2 r^4 \left(\frac{d\bar{\omega}}{dr}\right)^2 - \frac{1}{3G}r^3 \frac{dj^2}{dr}\bar{\omega}^2, \tag{6a}$$

$$\frac{dp_0^*}{dr} = -\frac{Gm_0(1+8\pi G r^2 p)}{r^2\left(1-\frac{2GM}{r}\right)^2} - \frac{4\pi G p_0^*(\epsilon+p)r}{\left(1-\frac{2GM}{r}\right)}$$
$$+ \frac{1}{12}\frac{r^3 j^2}{\left(1-\frac{2GM}{r}\right)}\left(\frac{d\bar{\omega}}{dr}\right)^2 + \frac{1}{3}\frac{d}{dr}\left(\frac{r^2 j^2 \bar{\omega}^2}{1-\frac{2GM}{r}}\right), \tag{6b}$$

where,

$$j = e^{-\nu_0/2}\left(1-\frac{2GM}{r}\right)^{\frac{1}{2}}.$$

Further, the expression for the angular velocity for the fluid relative to the local inertial frame, $\bar{\omega}$ is obtained from the equation

$$\frac{1}{r^4}\frac{d}{dr}\left(r^4 j \frac{d\bar{\omega}}{dr}\right) + \frac{4}{r}\frac{dj}{dr}\bar{\omega} = 0 \tag{6c}$$

The metric perturbation function $h_0$ is calculated using the algebraic relation obtained from hydrodynamic equation given as

$$h_0 = -p_0^* + \frac{1}{3}r^2 e^{-\nu_0}\bar{\omega}^2 + h_{0c}. \tag{7a}$$

Outside the star, $m_0$ and $h_0$ have the forms [3]

$$m_0 = \delta M - \frac{GJ^2}{r^3}, \tag{7b}$$

$$h_0 = -\frac{G\delta M}{r-2GM} - \frac{G^2 J^2}{r^3(r-2GM)}, \tag{7c}$$

where, $J$ is the angular momentum of the star given as

$$J = \frac{1}{6}\frac{R^4}{G}\left(\frac{d\omega}{dr}\right)_{r=R}.$$

The constant $h_{0c}$ in Eq. (7a) is obtained by matching the solution for $h_0$ at the boundary.

# III. ELECTROMAGNETIC FIELD STRUCTURE IN THE INTERIOR OF THE STAR

Compatible with an axisymmetric and stationary system, one can write down the Maxwell's equations in a curved background specified by Hartle metric as given in Eq.(2). The nonzero components of the electromagnetic fields satisfy the equations [9]

$$\frac{\partial}{\partial r}\left[e^{\psi+\mu}E_r\right] + \frac{\partial}{\partial \theta}\left[e^{\psi+\lambda}E_\theta\right] = -e^{\nu+\lambda+\psi+\mu}J^t, \tag{8a}$$

$$\frac{\partial}{\partial r}\left[\omega e^{\psi+\mu}E_r\right] + \frac{\partial}{\partial \theta}\left[\omega e^{\psi+\lambda}E_\theta\right] - \frac{\partial}{\partial r}\left[e^{\nu+\mu}B_\theta\right] + \frac{\partial}{\partial \theta}\left[e^{\nu+\lambda}B_r\right] = -e^{\nu+\lambda+\psi+\mu}J^\phi, \tag{8b}$$

$$\frac{\partial}{\partial r}\left[e^{\psi+\mu}B_r\right] + \frac{\partial}{\partial \theta}\left[e^{\psi+\lambda}B_\theta\right] = 0, \tag{8c}$$

$$-\frac{\partial}{\partial r}\left[e^{\nu+\mu}E_\theta\right] + \frac{\partial}{\partial \theta}\left[e^{\nu+\lambda}E_r\right] - \frac{\partial}{\partial r}\left[\omega e^{\psi+\mu}B_r\right] - \frac{\partial}{\partial \theta}\left[\omega e^{\psi+\lambda}B_\theta\right] = 0. \tag{8d}$$

To consider the interior electromagnetic field structure, we shall assume the interior of the star to be a perfect conductor. Hence the electric field inside the star is zero and, the magnetic field components as modified due to rotation upto monopole approximation for the metric are given as

$$\frac{\partial}{\partial r}[re^{\nu_0/2}(1+h_0)B_\theta] \\ - \frac{\partial}{\partial \theta}\left[e^{\nu_0/2}\left(1-\frac{2GM}{r}\right)^{-1/2}\left(1+h_0+\frac{Gm_0}{r-2GM}\right)B_r\right] = 0 \tag{9a}$$

$$\frac{\partial}{\partial r}[r^2 \sin\theta B_r] + \frac{\partial}{\partial \theta}\left[r\sin\theta \left(1-\frac{2GM}{r}\right)^{-1/2}\left(1+\frac{Gm_0}{r-2GM}\right)B_\theta\right] = 0 \tag{9b}$$

Assuming the magnetic field to be dipolar in nature, the field components can be written as

$$B_r = f(r)\cos\theta \quad B_\theta = g(r)\sin\theta, \tag{10}$$

and the equations (9a) and (9b) reduce to the coupled ordinary differential equations given as

$$r(1+h_0)g'(r) + \left[(1+h_0)\left(1+\frac{r}{2}\nu_0'(r)\right) + rh_0'(r)\right]$$
$$+ \left(1 - \frac{2GM}{r}\right)^{-1/2}\left(1 + h_0 + \frac{Gm_0}{r-2GM}\right)f(r) = 0 \tag{11a}$$

$$rf'(r) + 2f + 2\left(1 - \frac{2GM}{r}\right)^{-1/2}\left(1 + \frac{Gm_0}{r-2GM}\right)g(r) = 0 \tag{11b}$$

The boundary conditions for these two coupled differential equations are found by matching the interior solutions with the exterior solutions at the boundary and are given as [9,10]

$$f(R) = f_0(R) + \delta f(R), \tag{12a}$$

$$g(R) = g_0(R) + \delta g(R), \tag{12b}$$

where, $f_0(R)$ and $g_0(R)$ are the solutions corresponding to the nonrotating star [10] given as

$$f_0(R) = -\frac{3\mu}{4G^3M_0^3}\left[\ln\left(1 - \frac{2GM_0}{R}\right) + \frac{2GM_0}{R}\left(1 + \frac{GM_0}{R}\right)\right], \tag{13a}$$

$$g_0(R) = \frac{3\mu}{4G^2M_0^2 R}\left[\left(1 - \frac{2GM_0}{R}\right)^{-1} + \frac{R}{GM_0}\ln\left(1 - \frac{2GM_0}{R}\right) + 1\right]\left(1 - \frac{2GM_0}{R}\right)^{1/2}. \tag{13b}$$

with, $M_0 = M(R)$ is the mass of the nonrotating star obtained by solving the TOV equations (5a) and (5b). We now substitute equations (12) and (13) in equation (11) and expand in powers of $GM_0/R$ retaining terms upto monopole perturbation. Comparing the coefficients, we obtain the corrections to the nonrotating solutions in the monopole approximation as

$$\delta f(R) = \frac{12\mu G}{R^4}\left[\delta M\left(\frac{1}{4} + \frac{4}{5}\frac{GM_0}{R}\right) - \frac{GJ^2}{R^3}\left(\frac{1}{14} + \frac{3}{10}\frac{GM_0}{R}\right)\right] \tag{14a}$$

$$\delta g(R) = \frac{\mu G}{R^4}\left[\delta M\left(2 + \frac{37}{5}\frac{GM_0}{R}\right) - \frac{GJ^2}{R^3}\left(\frac{8}{7} + \frac{163}{35}\frac{GM_0}{R}\right)\right] \tag{14b}$$

## IV. EQUATIONS OF STATE AND QUARK HADRON PHASE TRANSITION

The interior of the star has density which is much higher than the normal nuclear densities. Unfortunately no single description of the equation of state will suffice to describe the whole density range from the center to the surface of the star. The main problem in determining the equations of state (EOS) at high densities lies in the uncertainity of the interactions operative and the procedure to calculate the same. The scope of the present work will be limited to take different equations of state and their effects on different stellar properties. We use here five neutron matter EOS and the bag model EOS for quark matter. The neutron matter EOS are (i) Wiringa's nonrelativistic EOS including the three body forces [11]; (ii) Walecka's mean field theory EOS obtained with nucleons interacting with scalar and vector bosons [12]; (iii) a field theoretical EOS given for high density matter given by Sahu et al, assuming the composition to be neutron rich matter in beta equilibrium based on the chiral sigma model [13]; (iv) Bethe-Johnson EOS based on a nonrelativistic constrained variational principle with a modified Reid soft core potential [14]; and (v) Pandharipande's hyperonic matter EOS based on many body variational theory [15].

For the densities below the nuclear density, the stellar matter is no longer determined by the nucleon nucleon interaction only and we shall use for the subnuclear density region, the EOS given by Negele and Vautherin [16] for densities of $10^{14}$ to $5 \times 10^{10}$ gm cm$^{-3}$, of Baym, Pethick and Sutherland [17] for densities of $5 \times 10^{10}$ gm cm$^{-3}$ till $\approx 10^3$ gm cm$^{-3}$ and of Feynman et al [18] for densities below $10^3$ gm cm$^{-3}$. In Fig. 1, we plot the equations of state for different models that we shall be using to study the neutron star structure. An EOS is said to be stiffer if the pressure is higher for a given energy density. However, as can be seen from fig. 1 that none of the EOS is stiffer as compared to another in the whole range of the energy density considered. Thus the stiffness of an EOS can be defined in an averaged sense from the value of the maximum stable mass of the neutron star using the particular EOS. An EOS is said to be stiffer on an average if it has higher value for $M_{max}$. In our calculations, however, we shall be referring to the stiff or soft nature of the EOS corresponding to the central energy density chosen for the star.

## A. Hybrid stars

Here we shall consider the case of a hybrid star i.e. a compact star with a quark matter core and a neutron matter crust. We shall be following the simplest approach of Ref [19] where one uses Bethe–Johnson EOS for neutron matter and bag model EOS for the quark matter phase. Bethe–Johnson EOS is based on a nonrelativistic constrained variational principle [14] with a modified 'Reid' soft core potential [20] as input and reads [21]:

$$\epsilon(n) = n[940.6 + 247.5n^{1.48}] \quad p(n) = 366.3n^{2.48}, \tag{15}$$

where $\epsilon(n)$ is the energy density, $p(n)$ the pressure in MeV/fm$^3$ as a function of baryon number density, $n$ in fm$^{-3}$. The quark gluon plasma phase is taken into account through a bag model EOS given as

$$p = \frac{1}{3}(\epsilon - 4B), \tag{16}$$

with the energy density of the QGP as that of a mixture of gluons and massless u-, d- and s- quarks at zero temperature given as

$$\epsilon(\mu_B) = \frac{1}{36\pi^2}(1 - \frac{2}{\pi}\alpha_s)\mu_B^4 + B. \tag{17}$$

In the above, $\mu_B$ is the baryon chemical potential, $\alpha_s$ is the strong coupling constant and $B$ is the bag constant. As in Ref. [19], we shall also take the strange quark mass to be zero. A finite strange mass $m_s$ will slightly decrease the pressure, which is inappreciable as compared to the role played by $B$ and $\alpha_s$ [19,22]. Further, since we take the mass of the strange quarks to be zero, the matter is electrically neutral without any further need for electrons [23]. Since we have only one chemical potential $\mu_B$, the phase boundary between hadronic and the QGP phase is determined by the Gibbs criteria as given by

$$p_{hadron}(\mu_{cr}) = p_{qgp}(\mu_{cr}) \tag{18}$$

where $\mu_{cr}$ is the critical baryon chemical potential.

In Fig. 2, we consider such a phase transition construction for the bag model EOS for quark matter and the Bethe-Johnson EOS for the neutron matter. We have taken the bag constant as $B^{1/4}$=190 MeV and the strong coupling constant $\alpha_s$=0.4. This leads to a critical chemical potential to be 1600 MeV and critical pressure, 407 MeV/fm$^3$. A reduction of B leads to a decrease in the critical chemical potential.

## V. RESULTS AND DISCUSSIONS

In order to construct a stellar model we proceed as follows. With a given central energy density as input, we integrate the coupled TOV equations (5a) and (5b) radially outwards from the center, using an equation of state until the pressure vanishes which defines the surface of the nonrotating star. This gives the mass ($M_0$) and radius ($R$) of the star as well as the pressure and mass profiles (p(r), M(r) respectively) to be used for later computations. The metric function $\nu_0(r)$ inside the star is then calculated by integrating Eq. (5c) from the surface to the center with the initial condition as

$$\nu_0(R) = ln\left(1 - \frac{2GM_0}{R}\right)$$

We next proceed to calculate the rotational perturbations by integrating equations (6a), (6b), (6c) from center to the surface for a given angular velocity at the center $\bar{\omega}_c$ with the initial conditions [3]

$$\frac{d\bar{\omega}}{dr}|_{r=0} = 0, \quad m_0(0) = 0 = p_0^*(0),$$

using the mass and pressure profiles obtained from solving the TOV equations. The second initial condition obviously implies that the central energy density of the rotating star is same as that of the non-rotating star [3].

With the monopole corrections calculated as above, we then proceed to integrate the dipole magnetic field configuration functions $f(r)$ and $g(r)$ given in eq.s (11b) and (11a) along with the respective boundary conditions given in Eq.(12a) and (12b).

The results with different equations of states are now given as follows.

We have plotted in fig. 3a the energy density profiles for stars with different neutron matter equations of state choosing the central energy density of the star as $\epsilon_c$=1600 MeV/fm$^3$. The drop in the profile near the surface corresponds to the crust of the star. The thickness of the crust increases with the softness of the EOS used to study the star [24]. The energy density profile is plotted for a pure quark star and a star with a quark core and neutron matter crust in fig. 3b, taking the central energy density as 3200 MeV/fm$^3$. The discontinuity in the energy density profile for the hybrid star at radius around 3.12 kms corresponds to the phase transition from quark matter to neutron matter. As the quark matter EOS is softer, it leads to a smaller mass and radius of the quark star.

For considering the rotational properties of the star, for all the EOS we take the rotating star to have the critical value for the angular velocity as given approximately by $\Omega_{cr} = \sqrt{\frac{GM}{R^3}}$. For central energy density of 1600MeV/fm$^3$, the values of the critical angular velocity are 1.758,1.551,1.524, 1.254 and 1.667 in units of $10^4$ sec$^{-1}$ for Wiringa, Walecka, Bethe-Johnson, Sahu et al and Pandharipande EOS respectively. We plot $d\bar{\omega}/dr$ as a function of the radial distance from the centre of the star with the neutron matter equations of state in fig.4a and for quark and hybrid stars in fig.4b. It starts with zero at the centre as given by the boundary condition and increases to a maximum and then decreases for all EOS except for quark matter EOS. The absence of a peak in the case of quark star is due to the fact that there is no crust for such stars and the energy density is finite at the surface even if the pressure vanishes, because of the nonzero nature of the bag constant. The height of the peak and its location depend both on the crustal thickness as well as the value of the critical angular velocity, $\Omega_{cr}$. As has been shown by Datta et al [24], the crustal thickness is larger for softer equation of state. The height of the peak increases with the stiffness of the EOS and increase in the value of $\Omega_{cr}$. Wiringa EOS which is the stiffest (for the $\epsilon_c$ considered) and has the highest value for $\Omega_{cr}$, exhibits the maximum peak. For the EOS given by Sahu et al, which though is not the softest EOS among those we consider, the star has the minimum value for $\Omega_{cr}$, which reflects in the peak being minimum. $\Omega_{cr}$ for the Walecka model is less than that for Pandharipande EOS, but is stiffer than the latter. These two effects compete with each other to make the peaks almost of the same height, where as the position of the peak from the surface increases with the crustal thickness as well as with $\Omega_{cr}$.

### A. Magnetic field topology

We then consider the field lines for different configurations and study their topology. In Fig. 5, for the central energy density, $\epsilon_c$=1600 MeV/fm$^3$, we have plotted the magnetic field lines for different EOS- (a) Bethe-Johnson, (b) Sahu et al, (c) Walecka and (d) Wiringa EOS. The corresponding values for the gravitational potential (GM$_0$/R) are 0.27, 0.284, 0.31 and 0.33 respectively. The difference in the field line densities seem to indicate that with the equation of state getting stiffer, the field line density increases indicating the necessity of higher fields to bind the matter configuration. This is consistent with the magnetohydrodynamics assumption of taking into account the concept of frozen in field lines.

We plot in fig.6 the magnetic field strength (for $\theta = \pi/4$) as a function of the radial distance in the interior of the star. We have chosen the value for $\mu$ approximately as $1.73 \times 10^{33}$ kms, corresponding to an equatorial magnetic field strength of around $10^{12}$ Gauss for a star of radius $\approx 10$ kms. As can be seen from the graph, the field strength at around a distance of 0.1 kms from the center is increased by about six orders of magnitude of the same at the surface for all the EOS.

Rotation also modifies the field strength. We next plot the percentage change in the field strength (for $\theta = \pi/4$) as compared to non rotating configurations for different EOS. As is clear from the Fig 7a, this increase as compared to the nonrotating case is more for the stiffer EOS. The field strength can get increased up to $\approx 33\%$ due to rotation as one approaches towards the core for Wiringa EOS, where as near the surface the increment is up to about 10% only.

We thus have considered here the effect of different EOS on the gross structural properties of the rotating neutron stars. We have also considered the case of hybrid stars with a quark matter core and neutron matter crust which have larger mass as compared to the pure quark star. Because of the quark hadron phase transition inside such hybrid stars, there is a discontinuity in the energy density profile $\epsilon(r)$ as well as in the rotational perturbations $m_0(r)$, $p_0(r)$, $\bar{\omega}(r)$ and $d\bar{\omega}/dr$. However, the discontinuity is more prominent in the energy density profile as compared to the rotational perturbation functions. We have also calculated the dipole magnetic field structure inside such compact objects. Stiffer EOS yield more modification in the magnetic field strength as compared to the nonrotating stars. This is a reflection of the fact that the rotational perturbations are more for the stiffer EOS. It may be noted that despite the fact that the EOS for Walecka model as well as for Sahu et al are in the unstable region for the central energy density chosen for the present calculations, the stellar properties as considered here show a consistent pattern. However, it is seen that the quark hadron phase transition does not show any discontinuity in the magnetic field strength for the hybrid stars.

To summarize, we have studied rotating compact stars endowed with magnetic fields and have discussed some of their gross properties for various EOS including the possibility of a hybrid star. However, in the present work, we have not included the effect of magnetic fields on the EOS. Such effects with a constant magnetic fields have been shown to soften the EOS [25] leading to a less massive star. For a more realistic dipole structure of the magnetic field, one does not yet know the EOS.

**Figures Captions**

Fig. 1. Plot of energy density ($\epsilon$) vs. pressure (p) each in units of MeV/fm$^3$.

Fig. 2. The pressure in MeV/fm$^3$ is plotted as function of baryon chemical potential, $\mu_B$. The solid line corresponds to the bag model EOS and the dashed line to the Bethe Johnson equation of state.

Fig. 3a. The energy density in MeV/fm$^3$ is plotted as function of the radial distance from the centre in kms with the neutron matter EOS.

Fig. 3b. The energy density in MeV/fm$^3$ is plotted as function of the radial distance from the centre in kms for the quark star and the hybrid star.

Fig. 4a. $\dfrac{d\bar{\omega}}{dr}$ in $10^4$/sec/kms vs the radial distance, r in kms with the neutron matter EOS.

Fig. 4b. $\dfrac{d\bar{\omega}}{dr}$ in $10^4$/sec/kms vs the radial distance, r in kms for the quark star and hybrid star.

Fig. 5. Field lines are plotted in X-Z plane for EOS given by Bethe-Johnson, Sahu et al, Walecka and Wiringa in (a), (b), (c) and (d) respectively.

Fig. 6. The logarithm of the magnetic field strength, B ($\theta = \pi/4$) in Gauss in the interior of the star.

Fig. 7a. The percentage increase in the field strength as a function of the radial distance in kilometers due to rotation of the star with neutron matter EOS.

Fig. 7b. The percentage increase in the field strength as a function of the radial distance in kilometers for the quark star and the hybrid star.

Fig.1

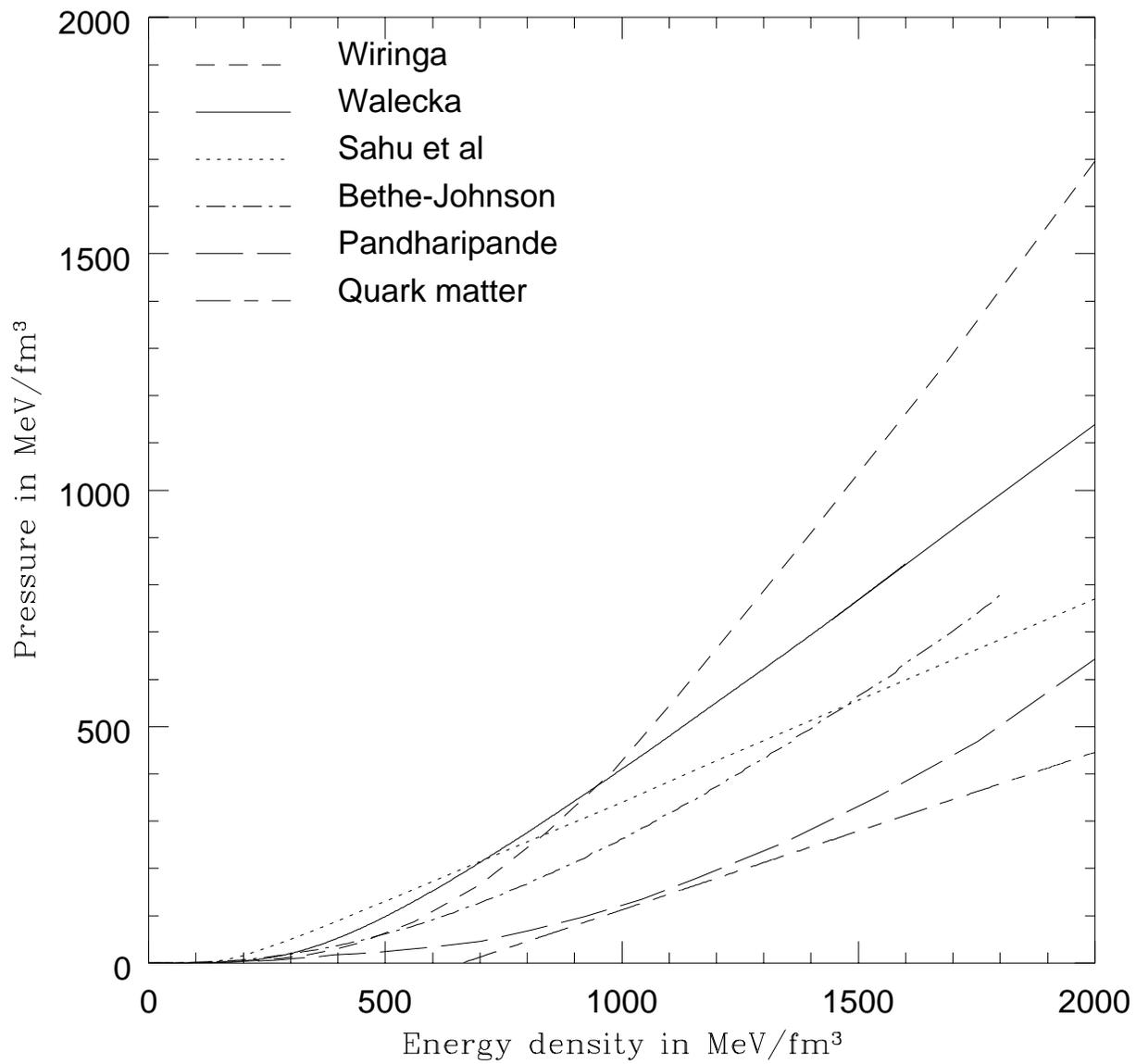

Fig. 2

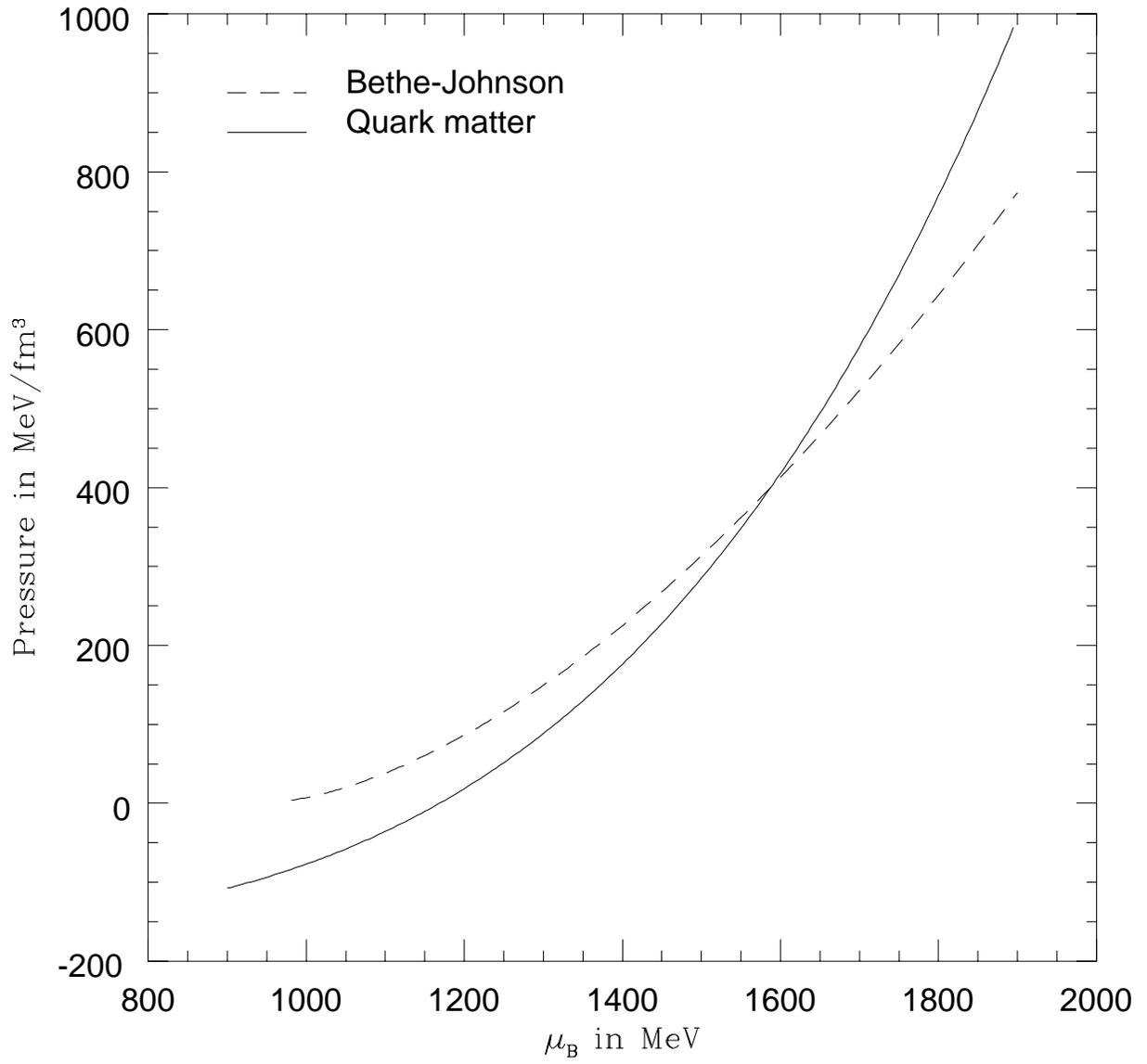

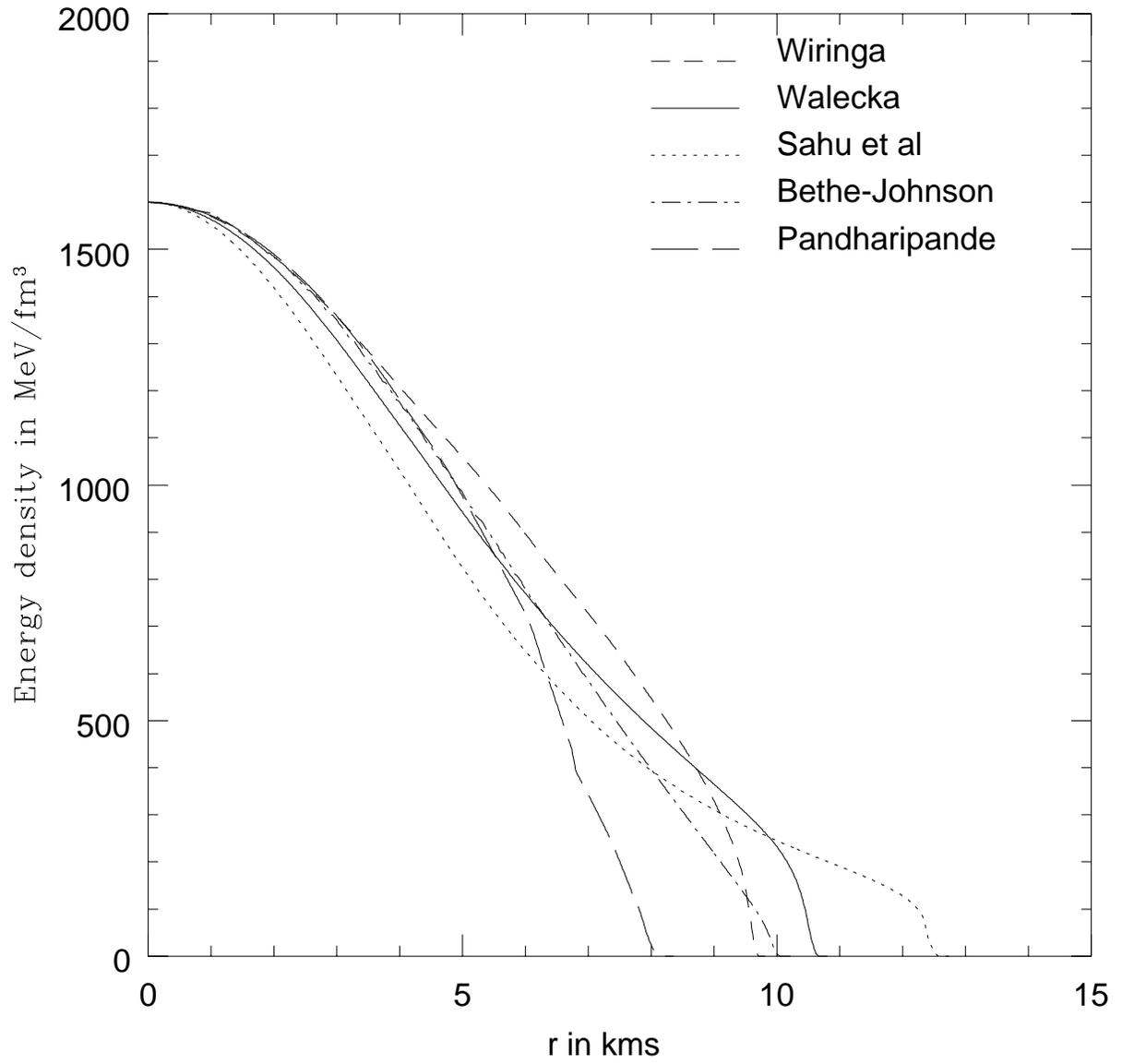

Fig. 3a

Fig. 3b

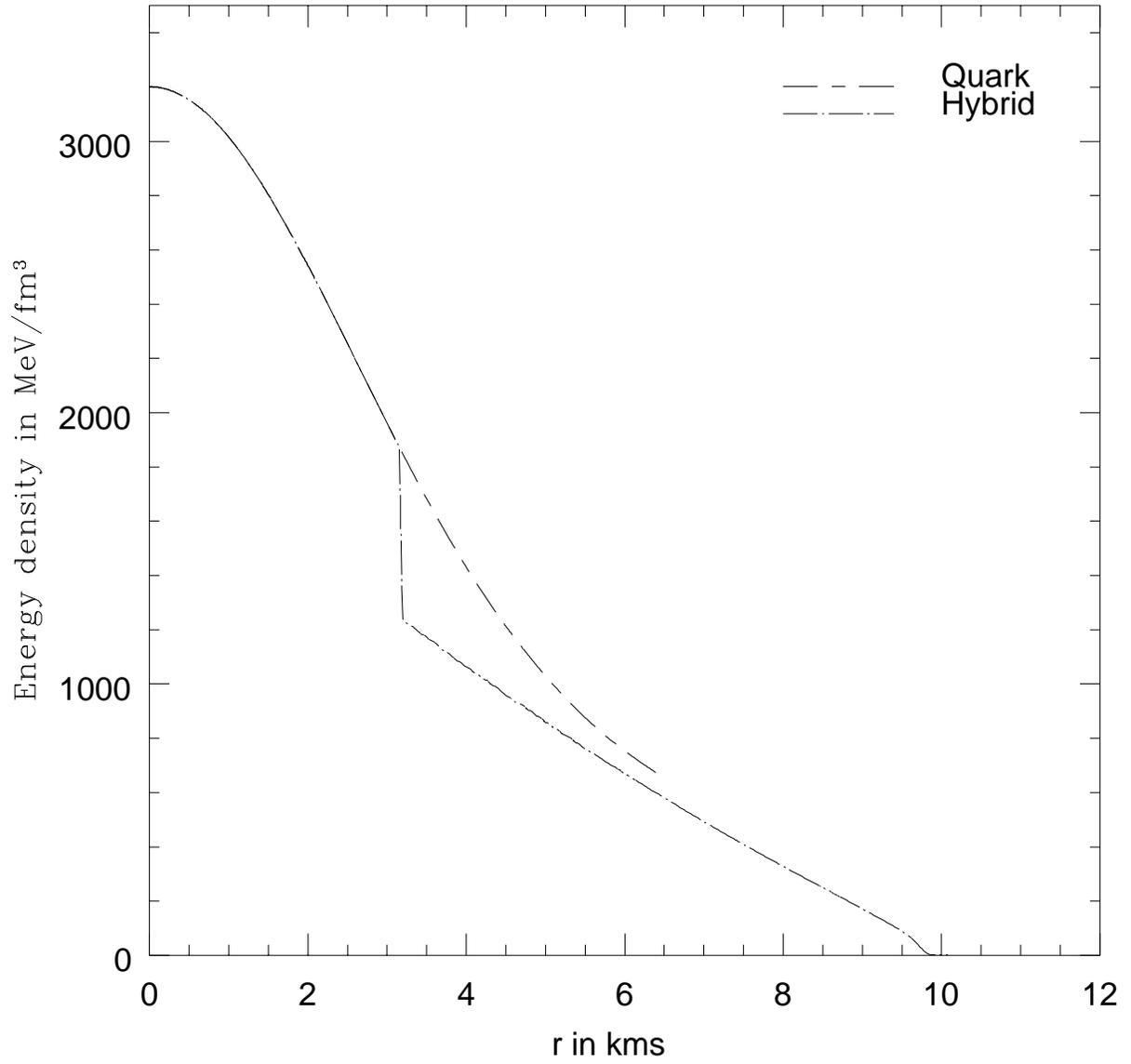

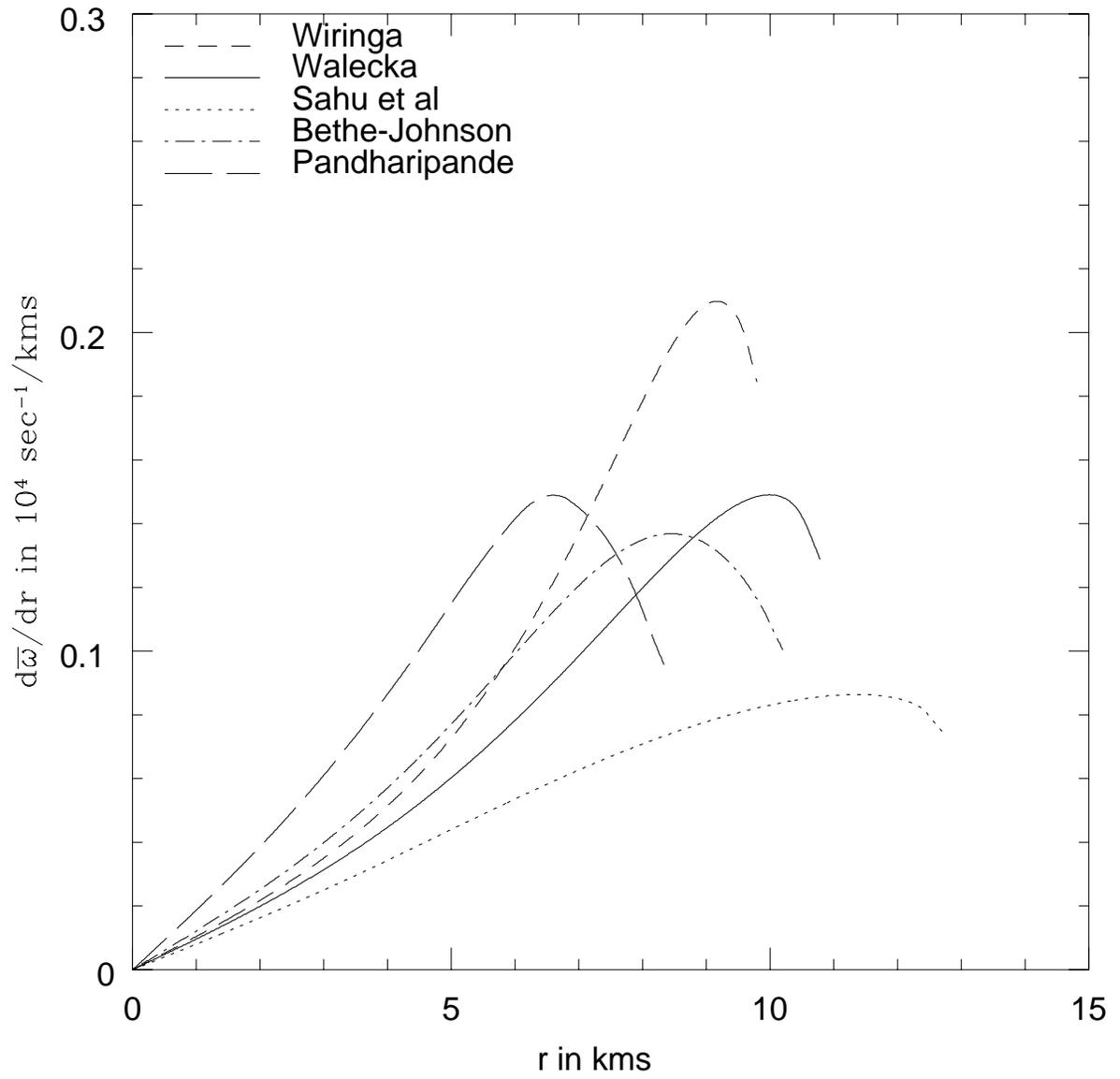

Fig. 4a

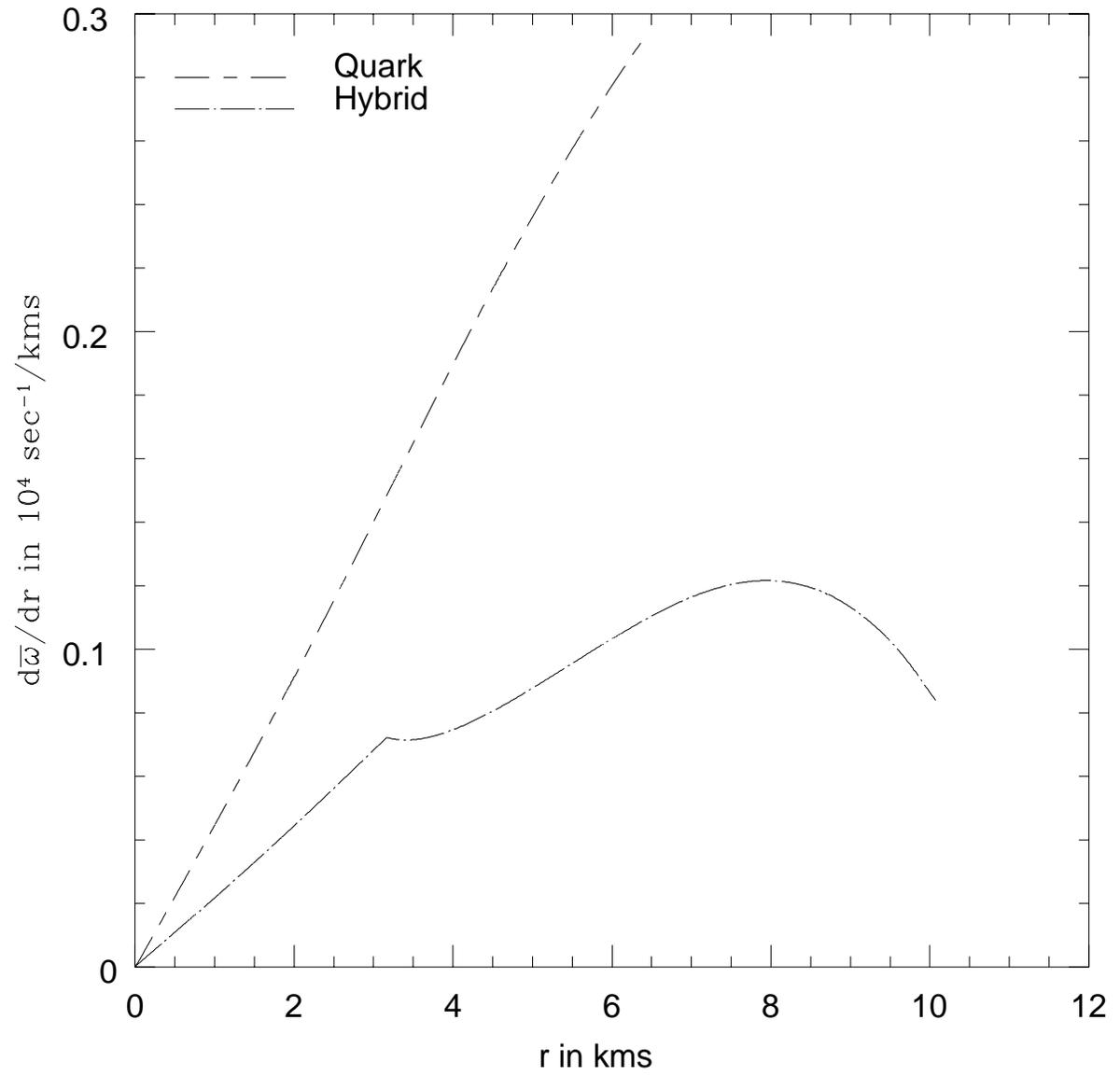

Fig. 4b

Fig. 5

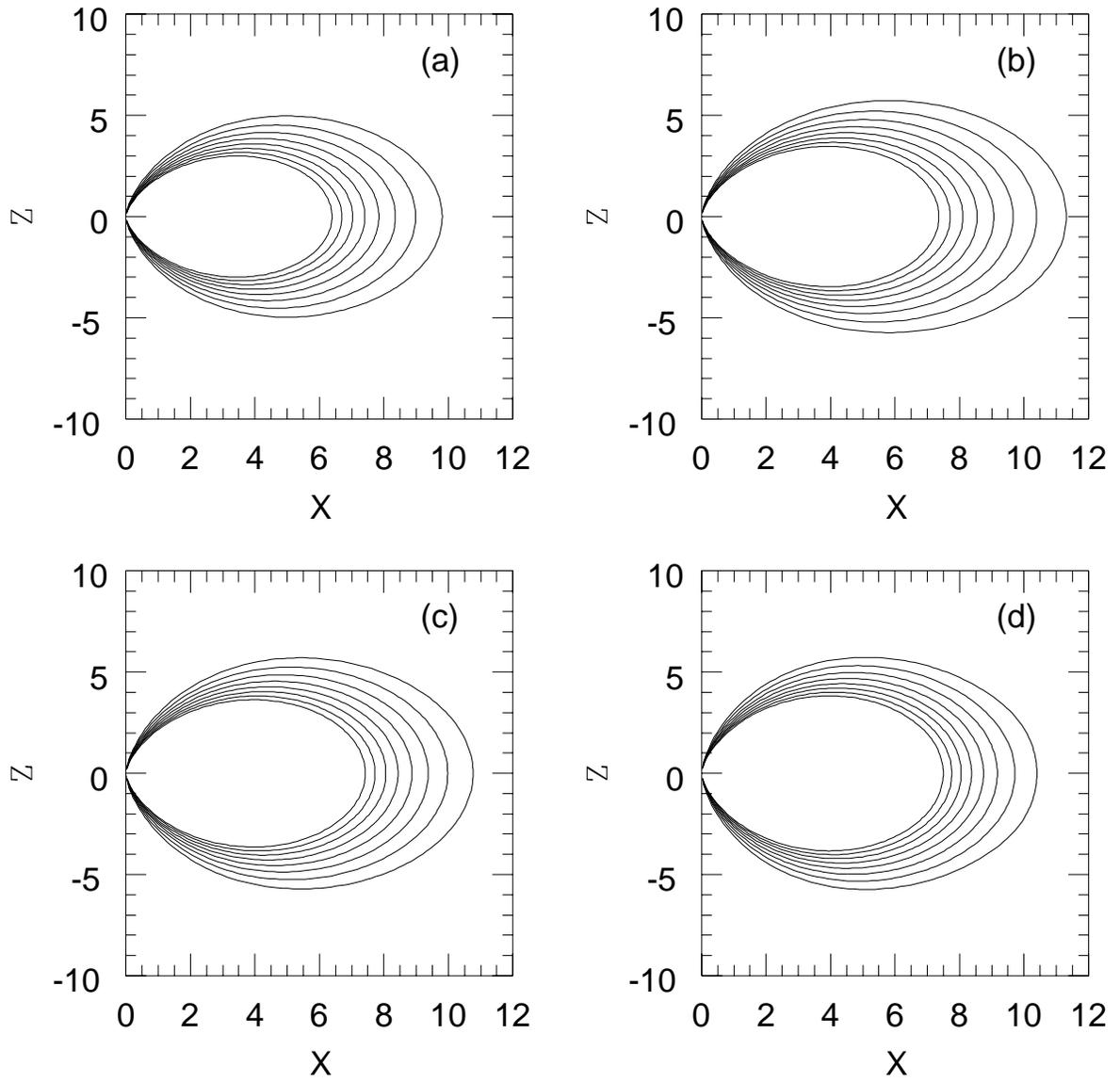

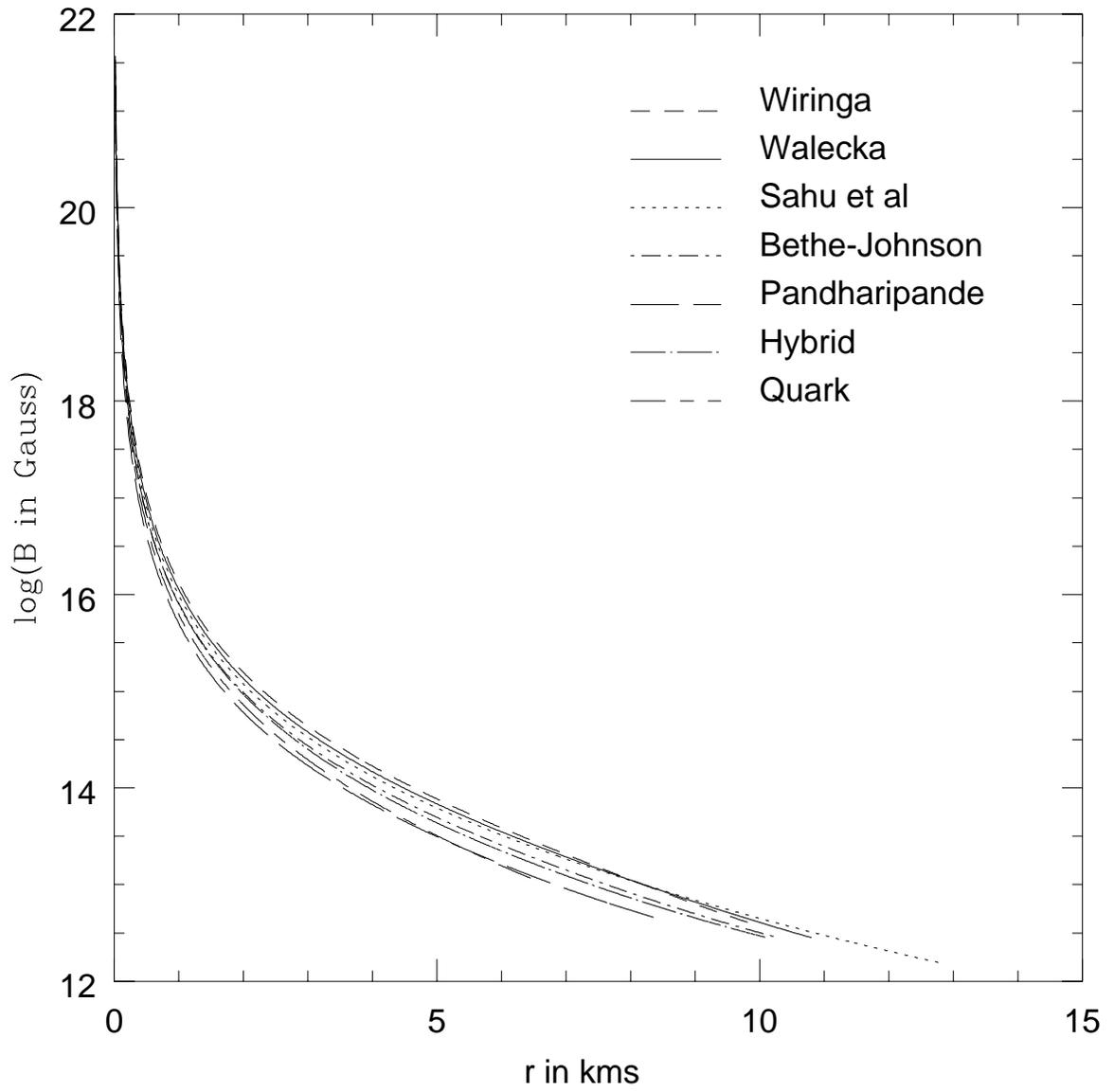

Fig. 6

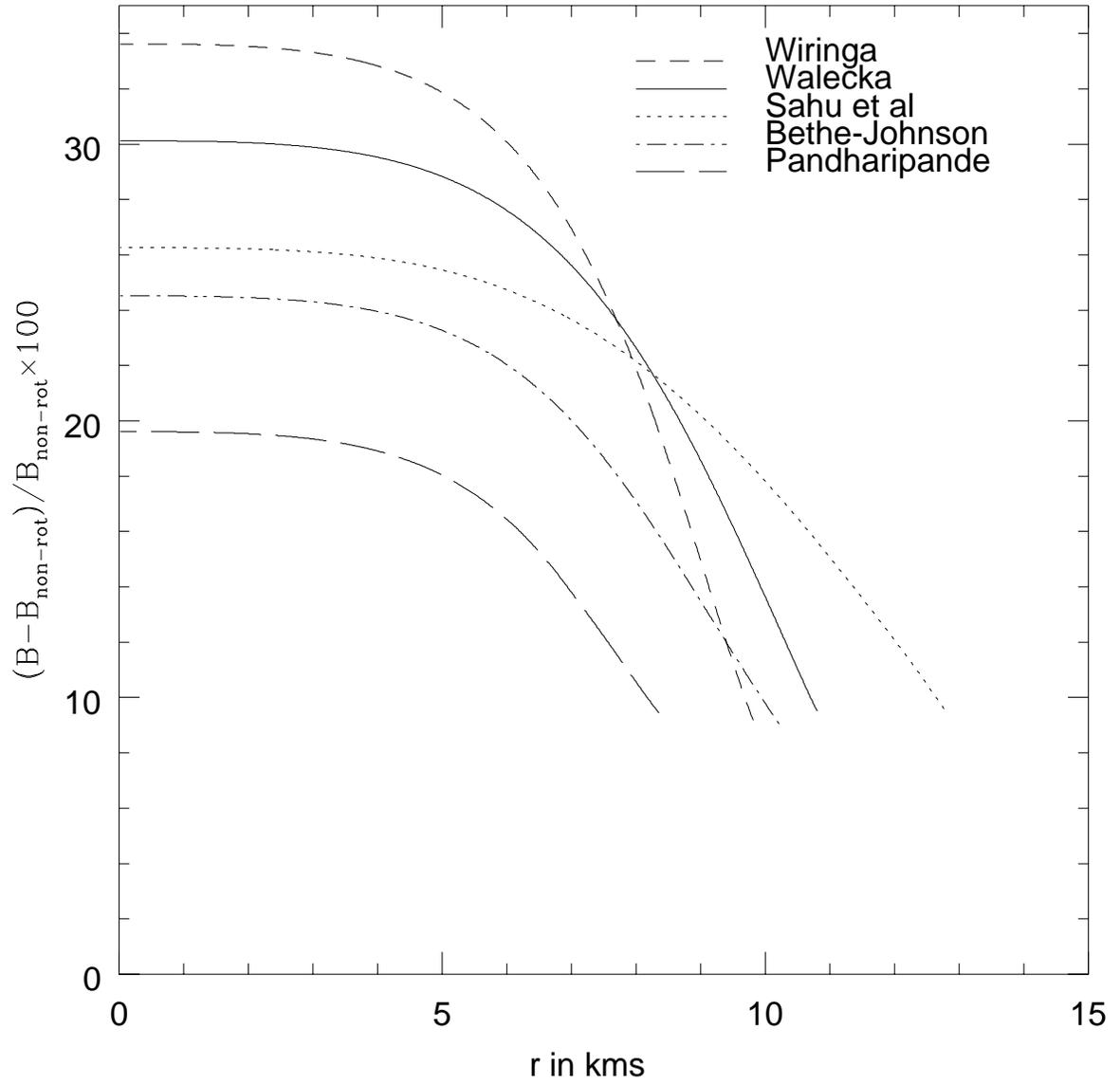

Fig. 7b

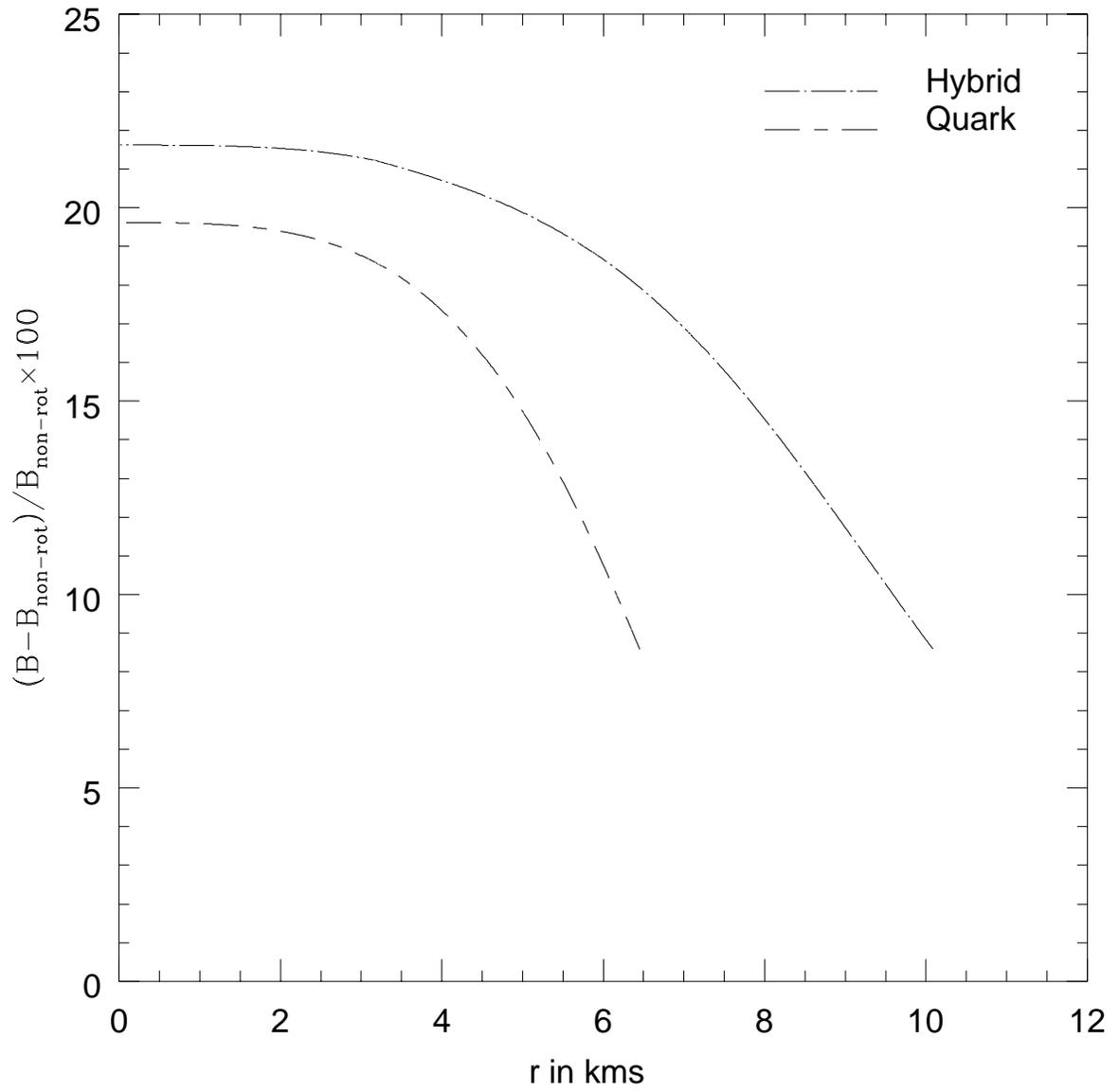